\documentclass[journal=nalefd,manuscript=letter,layout=twocolumn]{achemso}
\usepackage{color}
\usepackage{multirow,booktabs}
\usepackage{soul}
\usepackage[version=3]{mhchem}

\author{Nan Guo}
\author{Xiaoyang Chen}
\author{Tianlun Yu}
\author{Yu Fan}
\affiliation[Fudan University]
{Advanced Materials Laboratory, State Key Laboratory of Surface Physics, and Department of Physics, Fudan University, Shanghai 200433, China}
\author{Qinghua Zhang}
\affiliation
{Beijing National Laboratory for Condensed Matter Physics and Institute of Physics,
Chinese Academy of Sciences, Beijing 100190, China}
\author{Minyinan Lei}
\affiliation[Fudan University]
{Advanced Materials Laboratory, State Key Laboratory of Surface Physics, and Department of Physics, Fudan University, Shanghai 200433, China}
\author{Xiaofeng Xu}
\author{Xuetao Zhu}
\author{Jiandong Guo}
\author{Lin Gu}
\affiliation
{Beijing National Laboratory for Condensed Matter Physics and Institute of Physics,
Chinese Academy of Sciences, Beijing 100190, China}
\author{Haichao Xu}
\email{xuhaichao@fudan.edu.cn}
\affiliation[Fudan University]
{Advanced Materials Laboratory, State Key Laboratory of Surface Physics, and Department of Physics, Fudan University, Shanghai 200433, China}
\alsoaffiliation
{Shanghai Research Center for Quantum Sciences, Shanghai 201315, China}
\author{Rui Peng}
\email{pengrui@fudan.edu.cn}
\affiliation[Fudan University]
{Advanced Materials Laboratory, State Key Laboratory of Surface Physics, and Department of Physics, Fudan University, Shanghai 200433, China}
\alsoaffiliation
{Shanghai Research Center for Quantum Sciences, Shanghai 201315, China}
\author{Donglai Feng}
\email{dlfeng@ustc.edu.cn}
\affiliation
{National Synchrotron Radiation Laboratory and School of Nuclear Science and Technology, University of Science and Technology of China, Hefei 230026, China}
\alsoaffiliation
{Advanced Materials Laboratory, State Key Laboratory of Surface Physics, and Department of Physics, Fudan University, Shanghai 200433, China}
\alsoaffiliation
{New Cornerstone Science Laboratory, University of Science and Technology of China, Hefei 230026, China}
\alsoaffiliation
{Collaborative Innovation Center of Advanced Microstructures, Nanjing, 210093, China}

\title
  {Inferior interfacial superconductivity in 1 UC FeSe/SrVO$_3$/SrTiO$_3$ with screened interfacial electron-phonon coupling}

\abbreviations{EPC}

\begin{document}

\begin{abstract}
  Monolayer FeSe/TiO$_x$ and FeSe/FeO$_x$ interfaces exhibit significant superconductivity enhancement compared to bulk FeSe, with interfacial electron-phonon coupling (EPC) playing a crucial role. However, the reduced dimensionality in monolayer FeSe, which may drive superconducting fluctuations, complicates the understanding of the enhancement mechanisms. Here we construct a new superconducting interface: monolayer FeSe/SrVO$_3$/SrTiO$_3$, in which the itinerant electrons of highly metallic SrVO$_3$ films can screen all the high-energy Fuchs-Kliewer phonons, including those of SrTiO$_3$, making it the first FeSe/oxide system with screened interfacial EPC while maintaining the monolayer FeSe thickness. Despite comparable doping levels, the heavily electron-doped monolayer FeSe/SrVO$_3$ exhibits a lower pairing temperature ($T_\mathrm{g}$ $\sim$ 48 K) than FeSe/SrTiO$_3$ and FeSe/LaFeO$_3$. Our findings disentangle the contributions of interfacial EPC from dimensionality on enhancing $T_\mathrm{g}$ in FeSe/oxide interfaces, underscoring the importance of interfacial EPC in $T_\mathrm{g}$ enhancement. This FeSe/VO$_x$ interface also provides a platform for studying the interfacial superconductivity.
\end{abstract}

\section{Introduction}

Single unit-cell FeSe thin films grown on SrTiO$_3$ (001) substrate (1 UC FeSe/STO) have attracted intensive research interests over the past decades \cite{1st,ARPES,65K,LeeJJ,engineering}. 1 UC FeSe/STO hosts a superconducting pairing temperature ($T_\mathrm{g}$) as high as 65 K \cite{65K,65K_TSY} , which is close to the boiling point of liquid nitrogen, and significantly enhanced compared with the transition temperature ($T_\mathrm{c}$) of the bulk FeSe ( $\sim$ 8 K) \cite{bulk}. It is commonly recognized that the interfacial charge transfer from the oxygen vacancies in STO induces heavy electron doping in the top monolayer FeSe, and contributes to the enhancement of  $T_\mathrm{g}$ \cite{ARPES,65K,thickFeSe,chargetransfer,chargeimaging}. However, the superconducting $T_\mathrm{c}$ of various electron-doped FeSe \cite{KFeSe,LiFeOH,thickFeSe} is lower than the $T_\mathrm{g}$ of 1 UC FeSe/STO and there should be additional factors that further boost the superconducting pairing. In angle-resolved photoemission spectroscopy (ARPES) studies \cite{LeeJJ,LDH2012,LDH2015}, ``replica bands" were observed, suggesting that the electrons in FeSe couple with the Fuchs-Kliewer (FK) phonons in STO at small momentum transfer \cite{engineering,thickFeSe,TiO2}. Such interfacial electron-phonon coupling (EPC) is theoretically proposed to effectively enhance the superconducting pairing in heavy electron-doped FeSe/oxides \cite{LDHreview,ZSY,Rebec,Tang}. On the other hand, it is recently proposed that pairing strength is possibly enhanced by two-dimensionality \cite{2Dintercalate}. For example, in (TBA)$_x$FeSe, by intercalating organic ion layers between adjacent FeSe layers to push the system closer to the ideal two-dimensional limit, the superconducting gap is enhanced 
\cite{2Dintercalate}, and the $T_\mathrm{g}$ is observed at a higher temperature than the $T_\mathrm{c}$ \cite{2Dintercalate}, similar to the behavior in 1 UC FeSe/STO  \cite{2Dpseudogap}. It remains perplexing whether both interfacial EPC and two-dimensionality or one of the two factors contribute to the superconducting pairing enhancement in 1 UC FeSe/STO.

Numerous studies have focused on engineering the interfacial EPC to gain insights into its role in superconducting pairing. 
For instance, experiments on increasing the thickness of FeSe grown on SrTiO$_3$ from 1 unit cell to 20 unit cells and tuning the doping level with surface potassium dosing show that the decay of the superconducting gap correlates with the decay of interfacial EPC, emphasizing the importance of interfacial EPC in superconductivity \cite{ZWH,ZSY}. 
However, the varied FeSe thickness between 1 unit cell and 20 unit cells in FeSe/STO makes it challenging to isolate the effect of interfacial EPC from that of dimensionality. 
More recently, the interfacial EPC strength has also been successfully modulated in 1 UC FeSe/STO, revealing a linear relationship between the gap size and the interfacial EPC constant ($\lambda$)  \cite{SQ}.
However, 1 UC FeSe/STO in the limit of zero interfacial EPC could not be reached, and the role of dimensionality can not be distinguished.
To discern the effect of interfacial EPC and two-dimensionality, a promising approach is to completely screen the interfacial EPC while maintaining the monolayer thickness of FeSe.

To screen the FK phonons, we here insert highly metallic films between FeSe and STO, whose itinerant electrons provide strong screening effects. Note that the 93 meV FK phonons generate a strong dipolar field perpendicular to the interface, which could not be screened by two-dimensional electron gas \cite{FeSeLTO}. Metallic film with three-dimensional bulk metallicity should be used. Here we designed and obtained high-quality 1 UC FeSe/SrVO$_3$/Nb:SrTiO$_3$ (FeSe/SVO/STO) heterostructures. By systematic studies on the interfacial atomic structure, electronic structure, and phonon spectra, we demonstrate that 1 UC FeSe/SVO/STO is a high-quality interfacial superconductor with superconducting $T_\mathrm{g}$ of $\sim$ 48 K, in which the interaction between FeSe electrons and the FK phonons in oxides is fully screened.  Our experimental findings reveal a superconducting pairing temperature of 48 K in the absence of interfacial EPC, while the increase from 48 K to 65 K can be solely attributed to the effect of interfacial EPC. These results disentangle the contributions from dimensionality and interfacial EPC on the enhancement of the pairing temperature in 1 UC FeSe/oxides.

\section{Results and discussion}

\begin{figure*}[h!]
		\includegraphics[width=132mm]{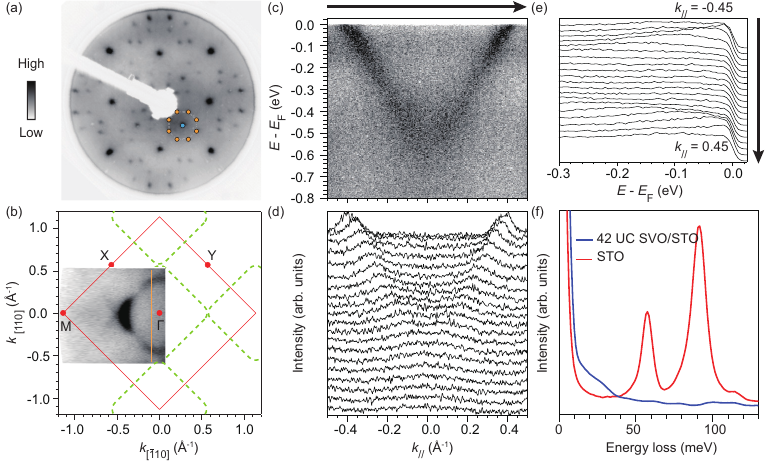}
		\caption{\textbf{Electronic structure of SVO/STO and screening of the FK phonons.} \textbf{a} Low-energy electron diffraction (LEED) patterns of 42 UC SVO/STO. Blue and orange filled circles represent the $\sqrt{2}\times\sqrt{2}$ R$45^{\circ}$ and $\sqrt{5}\times\sqrt{5}$ R$26.6^{\circ}$ surface reconstructions, respectively. \textbf{b} In-plane photoemission intensity map of 42 UC SVO/STO integrated over [$E_\mathrm{F}-100$ meV, $E_\mathrm{F}+100$ meV] using 21.2 eV photons. Green dashed lines schematically illustrate the Fermi surfaces of SVO/STO. \textbf{c} Photoemission spectra near the $\Gamma$ point as indicated by the orange line in panel \textbf{b}.  \textbf{d} The corresponding Momentum distribution curves (MDCs) taken at [$E_\mathrm{F}-0.8$ eV, $E_\mathrm{F}$] of the spectra in \textbf{c}. \textbf{e} Energy distribution curves (EDCs) along cut in (c). \textbf{f} Electron energy loss spectra of 42 UC SVO/STO and bare STO substrate.}
		\label{fig1}
		\end{figure*}

SVO is a perovskite-type transition-metal oxide whose lattice constant (3.8425 $\AA$) is comparable to that of STO (3.905 $\AA$) \cite{SVOstructure}. It exhibits metallic behavior when the thickness exceeds 5 UC \cite{SVOMIT, SVOmetallic}. Thick SVO films with VO$_2$ termination were epitaxially grown on the commercial Nb-doped STO (001) substrates at 750 $^{\circ}$C under an oxygen pressure of 1.9 $\times 10^{-7}$ mbar by molecular beam epitaxy (see details in Supplementary Note 1). The low energy electron diffraction (LEED) patterns of SVO films (Figure 1a) shows sharp diffraction spots, indicating a high-quality two-dimensional surface. Two distinct surface reconstructions, $\sqrt{5}\times\sqrt{5}$ R$26.6^{\circ}$ and $\sqrt{2}\times\sqrt{2}$ R$45^{\circ}$, are observed. The complex surface reconstruction does not affect the subsequent growth of tetragonal FeSe, which is in line with the previous report on the growth of FeSe on the reconstructed STO surfaces \cite{PR}.

The Fermi surface (FS) of thick SVO films, measured using 21.2 eV photons, mainly consists a rounded-square pocket around the $\Gamma$ point, and weak parallel-line spectral weight extending along $\Gamma$-X and $\Gamma$-Y directions (Figure 1b). The ARPES spectra along $\Gamma$-M direction (Figure 1c) and the corresponding momentum distribution curves (MDCs) (Figure 1d) show an electron-like dispersion which is contributed by the $d_{xy}$ orbital of vanadium according to previous reports \cite{SVOorbital,SVOARPES}. The parallel features could be contributed by the spectral weight from d$_{xz}$/d$_{yz}$ orbitals \cite{SVOorbital} with finite k$_z$ broadening or $\sqrt{2}\times\sqrt{2}$ R$45^{\circ}$ reconstruction-induced band folding \cite{SVOreconstruction,STOreconstruction}. The energy distribution curves (EDCs) display pronounced Fermi cut-offs at $E_\mathrm{F}$ (Figure 1e), demonstrating the metallic nature of the SVO films. Considering three cylindrical FSs of $d_{xy}, d_{yz}$ and $d_{xz}$ \cite{SVOorbital,SVOARPES}, the three-dimensional carrier density of SVO films estimated by the Luttinger theorem \cite{Luttinger} is approximately $2 \times 10^{22}$ /cm$^{3}$. Such a large density of itinerant electrons in thick SVO films is expected to provide strong screening effects to the dipolar field of FK phonons \cite{screen_length}. 

        \begin{figure*}[h!]
		\includegraphics[width=165mm]{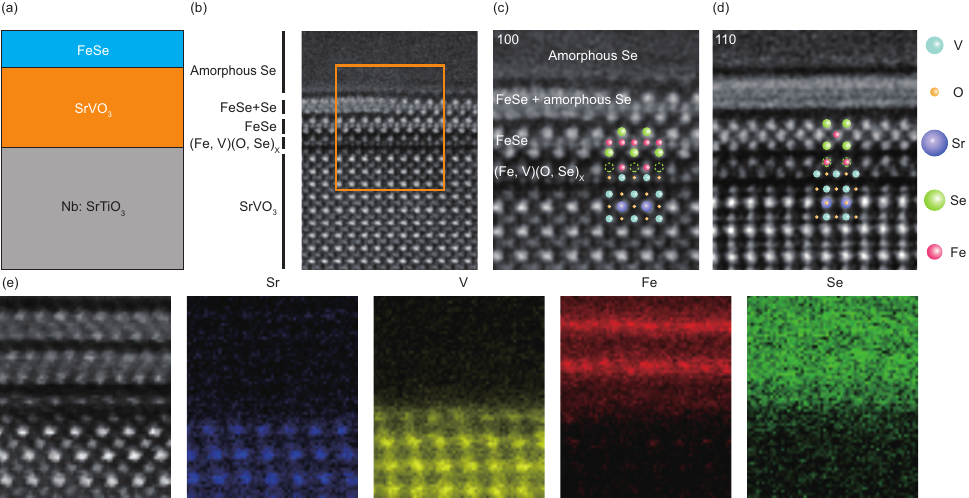}
		\caption{\textbf{Interfacial atomic structure of Se capped 1.5 UC FeSe/SVO/STO.} \textbf{a} Schematic illustration of the FeSe/SVO/STO heterostructure. \textbf{b} High angle annular dark field (HAADF) image of amorphous Se capped 1.5 UC FeSe/SVO/STO along the [100] direction of the SVO lattice. \textbf{c} Zoom-in HAADF image near the FeSe/SVO interface as indicated by the orange box in (b), and the side view of the corresponding atomic structure is overlaid. \textbf{d} The same as \textbf{c} but along [110] direction. \textbf{e} Element-resolved maps based on the EELS data at the Sr-$L_{2,3}$, V-$L_{2,3}$, Fe-$L_{2,3}$ and Se-$L_{2,3}$ absorption edges. The first image is the region where the measurement is applied.}
		\label{fig2}
		\end{figure*}
 
To assess the screening effects on interfacial phonons, high-resolution electron energy loss (HR-EELS) measurements were performed on SVO/STO to study the phonon spectra and compare them to those at the STO surface (Figure 1f). The HR-EELS spectra of STO surface show two prominent peaks at $\sim$ 60 meV and $\sim$ 92 meV, which correspond to the FK surface phonons of STO \cite{FKphonon,STOFK}. However, these two peaks disappear in SVO/STO. Given that the incident electrons are highly sensitive to the dipole electric field \cite{EELS,ZSY}, signals of the STO FK phonons should be detected if they could penetrate through the top SVO films\cite{ZSY,SYH}. Moreover, signals of the SVO phonons are also weak in the energy range of 80 $\sim$ 100 meV. Overall, the FK phonons either from SVO or STO in the SVO/STO heterostructure get screened by the mobile electrons in SVO.

FeSe films were grown on the thick SVO films subsequently at 490 $^{\circ}$C by co-evaporating Fe and Se and then post-annealed at 520 $^{\circ}$C for 3 hours (Figure 2a). The cross-sectional atomic structure of amorphous-Se-capped 1.5 UC FeSe/SVO/STO was resolved at the atomic level using scanning transmission electron microscopy (STEM) (Figure 2b-2d). The first layer of FeSe shows clear atomic structure and no intermixing with the capped Se.

An atomically-resolved intermediate layer between the SVO films and the first layer of FeSe is observed. By analyzing the elementary-resolved maps (Figure 2e), it was observed that the top layer of $\mathrm{VO_x}$ double layers \cite{doublelayer},  as well as some Fe and Se atoms diffusing downward, contribute to the formation of this intermediate layer

(see details in Supplementary Note 2).  The first layer of FeSe and the intermediate layer form a sharp interface, indicating that FeSe can be well grown on the VO$_x$-terminated SVO. Our results provide a new interfacial system of monolayer FeSe grown on non-TiO$_x$ terminated oxides after FeSe/MgO \cite{FeSeMgO}, FeSe/NdGaO$_3$ \cite{FeSeNGO} and FeSe/LaFeO$_3$ (FeSe/LFO) \cite{SYH}.

 \begin{figure*}[htbp]
    \includegraphics[width=170mm]{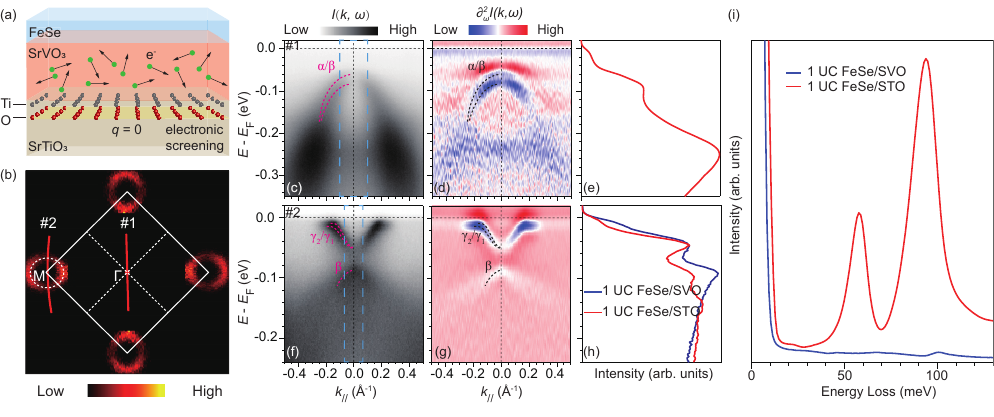}
    \caption{\textbf{Electronic structure of 1 UC FeSe/SVO/STO.} \textbf{a} Sketch of FeSe/SVO/STO heterostructures and the screening of phonons by itinerant electrons. \textbf{b} Photoemission intensity map of 1 UC FeSe/SVO/STO integrated over [$E_{F}-20$ meV, $E_{F}+20$ meV] using 21.2 eV photons. \textbf{c-e}  Photoemission spectra along cut \#1 in b (\textbf{c}), the second derivative of the photoemission intensity with respect to energy (\textbf{d}), EDC integrated over the momentum range indicated by the blue dashed rectangle in c (\textbf{e}). \textbf{f-h} The same as \textbf{c-e}, but along cut \#2 in panel \textbf{b}. The EDC of 1 UC FeSe/STO from ref. \cite{SQ} with the same integration range is appended in h. \textbf{i} Electron energy loss spectra of 1 UC FeSe/STO and 1 UC FeSe/SVO/STO. }
    \label{fig3}
    \end{figure*}

In-situ ARPES measurements were conducted on 1 UC FeSe/SVO/STO heterostructures to study the electronic structure. The FS of 1 UC FeSe/SVO is composed of two elliptical electron pockets located at the Brillouin zone corner, while no hole pocket at the Brillouin zone center (Figure 3b), indicating that the monolayer FeSe is heavily electron-doped by the interface, similar to 1 UC FeSe/STO and 1 UC FeSe/LFO. The photoemission spectra along high symmetric cuts \#1 (Figure 3c-e) exhibit two parabolic bands (noted as $\alpha$ and $\beta$) located near the $\Gamma$ point with their band top at $E-E_\mathrm{F} \sim -59$ meV. Along high symmetric cuts \#2 (Figure 3f-h), two-electron bands (noted as $\gamma_1$ and $\gamma_2$) are resolved around the M point with their band bottom at $E-E_\mathrm{F} \sim -51$ meV. According to Luttinger theorem \cite{Luttinger}, the doping level was estimated to be about 0.088 $e^{-}$/Fe based on the Fermi surface volume, slightly lower than the optimal doping (0.12 $e^-/\mathrm{Fe}$) of 1 UC FeSe/STO \cite{thickFeSe}, but almost identical to that of 1 UC FeSe/LFO (0.087 $e^-/\mathrm{Fe}$) \cite{SYH}, indicating that 1 UC FeSe/SVO and 1 UC FeSe/LFO should locate at the same position in the phase diagram when only considering the effect of electron doping \cite{thickFeSe}.

 \begin{figure*}[h!]
    \includegraphics[width=170mm]{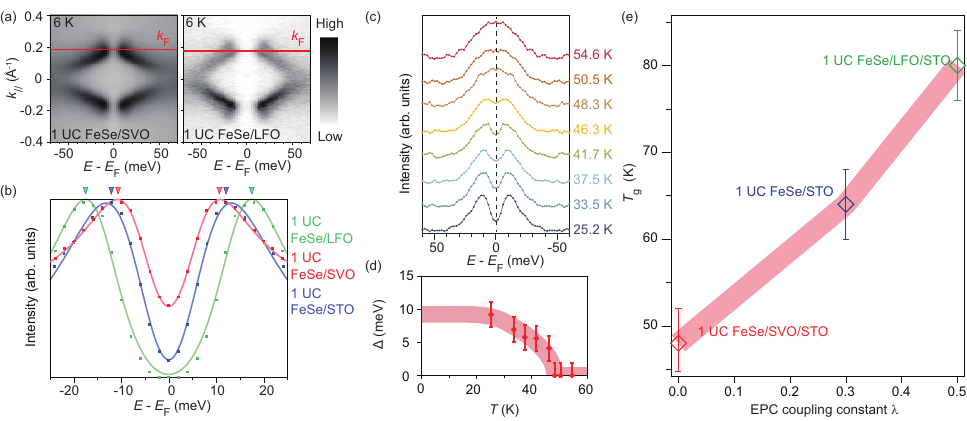}
    \caption{\textbf{Superconducting gap of 1 UC FeSe/SVO/STO.} \textbf{a} Symmetrized photoemission spectra with respect to the corresponding $E_\mathrm{F}$ at 6 K of 1 UC FeSe/LFO/STO and 1 UC FeSe/SVO/STO. \textbf{b} Symmetrized EDCs at the corresponding Fermi momentum $k_\mathrm{F}$ of 1 UC FeSe/SVO, 1 UC FeSe/STO \cite{SQ}, and 1 UC FeSe/LFO \cite{SYH}. The fittings to a superconducting function based on the simplified Bardeen-Cooper-Schrieffer (BCS) self-energy are appended. \textbf{c, d} Temperature dependence of the symmetrized EDC at the $k_\mathrm{F}$ (\textbf{c}), and the determined superconducting gap (\textbf{d}) of 1 UC FeSe/SVO/STO. The error bars of the gap are from the standard deviation of the fitting process and the measurement uncertainty. \textbf{e} Superconducting gap close temperatures ($T_\mathrm{g}$s) of 1 UC FeSe/SVO/STO, 1 UC FeSe/STO, and 1 UC FeSe/LFO/STO are plotted as a function of the interfacial EPC constant $\lambda$.}
    \label{fig4}
    \end{figure*}	

As the small-$q$ FK phonons of STO are screened by the itinerant electrons in thick SVO layers (Figure 3a), the replica bands, which are the signature of interfacial EPC \cite{LeeJJ,LDH2015,forwardscattering,WY2016,WY2017}, were expected to be absent. Consistently, in the ARPES spectra of 1 UC FeSe/SVO/STO, replica bands for all the main bands of monolayer FeSe ($\alpha$, $\beta$, $\gamma_1$, and $\gamma_2$)  are not observed  [Figs. 3c-h], in contrast with the ARPES spectra of 1 UC FeSe/STO and 1 UC FeSe/LFO, in which the prominent band replicas are observed at $\sim$ 100 meV, and $\sim$ 80 meV below the main bands, respectively \cite{LeeJJ, SQ,SYH}. Compared with the high-quality 1 UC FeSe/STO \cite{SQ}, the momentum-integrated EDCs near the M point show a comparable signal-to-noise ratio and similar main band intensity (Figure 3h), and the replica should be observed if exists. However, the lineshape at the replica band region is dramatically different, with no replica band intensity in 1uc FeSe/SVO. 
The analysis of the quasiparticle lifetime also suggests a similar scattering rate in 1 UC FeSe/SVO/STO.
 
Therefore, the absence of replica bands in 1 UC FeSe/SVO/STO is not an effect of the sample quality. 
It should be noted that the absence of replica bands is consistent across multiple ARPES measurements and analyses conducted on several 1 UC FeSe/SVO/STO samples. Consistent with ARPES results, the HR-EELS measurements show the significantly suppressed FK phonon peaks in 1 UC FeSe/SVO/STO in comparison with that of 1 UC FeSe/STO (Figure 3i). All these findings suggest that the dipole field generated by the FK phonons is incapable of penetrating through the thick SVO layers and the interfacial EPC is well-screened in 1 UC FeSe/SVO/STO.

    \begin{table*}[htp]
    \centering
    \begin{tabular}{cccc}
    \toprule[1pt]
    FeSe/oxides & doping level ($e^-$/Fe) & $\Delta_1$ (meV) & $\Gamma_1$ (meV) \\
    \midrule[1pt]
    1 UC FeSe/SVO & 0.088 & 9.6 & 21.7 \\
    \midrule[1pt]
    1 UC FeSe/STO & 0.12 & 12.1 & 20.8 \\
    \midrule[1pt]
    1 UC FeSe/LFO & 0.087 & 17 & 20 \\
    \bottomrule[1pt]
    \end{tabular}
    \caption{Summary of doping level, superconducting gap size $\Delta_1$, and quasiparticle lifetime $\Gamma_1$ of 1 UC FeSe/SVO, 1 UC FeSe/STO \cite{SQ}, and 1 UC FeSe/LFO \cite{SYH}.}
    \end{table*}

As shown in Fig.4a, the electron bands ($\gamma_2/\gamma_1$) exhibit an evident back bending near $E_\mathrm{F}$ at low temperature, which is a characteristic behavior of the Bogoliubov quasiparticle dispersion. With almost the same electron doping level, 1 UC FeSe/LFO has a significantly larger superconducting gap (see the right panel of Figure 4a).
To quantitatively compare the size of the superconducting gap, the EDC at Fermi momentum ($k_\mathrm{F}$) is fitted by the superconducting spectral function with the simplified BCS self-energy $\Sigma(\boldsymbol{k}, \omega)=-\mathrm{i} \Gamma_{1}+\Delta^{2} /\left[\omega+\epsilon(k)\right]$ \cite{gapfitting,BaFeAsP,SQ} (Figure 4b), where $\Delta$ is the superconducting gap, $\Gamma_1$ is the energy broadening caused by the quasiparticle scattering off impurities, and hence an indication of the quasiparticle lifetime. The three films show similar values of $\Gamma_1$, where $\Gamma_1$(FeSe/SVO) = 21.7~meV, $\Gamma_1$(FeSe/STO) = 20.8~meV, and $\Gamma_1$(FeSe/LFO) = 20~meV, suggesting similar sample quality. 
On the other hand, the superconducting gap size of  $\gamma_1$ band ($\Delta_1$) of FeSe/SVO ($\Delta_1$ = 9.6~meV) is significantly smaller than that of 1 UC FeSe/STO ($\Delta_1$=12.1 meV) and FeSe/LFO ($\Delta_1$=17 meV) \cite{SQ, SYH}. These values are summarized in Table 1 to visually compare these three systems.
Since the sample qualities are comparable,  the reduced gap size indicates a weaker pairing strength in 1 UC FeSe/SVO/STO.

Temperature-dependent measurements to determine the $T_\mathrm{g}$ were conducted on the same sample (Figure 4c). The gap gradually decreases with increasing temperature and eventually closes at $T_\mathrm{g}$ = 48 K, which is well fitted by the BCS formula (Figure 4d). 

~\\

\subsection{Discussion}
Monolayer FeSe grown on SVO/STO demonstrates a high-quality heterostructure with a well-ordered interfacial atomic structure. The quasiparticle scattering rate is similar to that in 1 UC FeSe/STO and FeSe/LFO. The well-screened interfacial EPC makes it a prototypical system to investigate the role of the interfacial EPC in FeSe/oxide interfacial systems at the 1 UC limit.
The gap-closing temperature $T_\mathrm{g}$, which represents the onset of the superconducting pairing, can be directly compared among the 1 UC FeSe/SVO, FeSe/STO, and FeSe/LFO as a function of interfacial EPC strength while preserving the identical thickness of FeSe.
A monotonic suppression of the $T_\mathrm{g}$ is observed as the interfacial EPC strength decreases from the 1 UC FeSe/LFO, 1 UC FeSe/STO to 1 UC FeSe/SVO (Figure 4e).
Therefore, our observations emphasize the crucial role of interfacial EPC in the enhancement of the superconducting pairing in 1 UC FeSe/oxide systems.

It is also noticeable that the $T_\mathrm{g}$ of 1 UC FeSe/SVO/STO is quite similar to the highest $T_\mathrm{g}$ of surface potassium-dosed multilayer FeSe, in spite of their different FeSe thicknesses, suggesting a minimal influence of dimensionality on the superconducting pairing in these systems. In regard to the higher $T_\mathrm{g}$ in organic molecules intercalated FeSe \cite{2Dintercalate}, the contribution of the coupling between FeSe electrons and certain vibrational modes of organic molecules may need to be investigated and taken into account.

Moreover, the discovery of the high-quality FeSe/VO$_x$ interface and its superconductivity provide a new playground for studying the FeSe/oxide interfaces. Perovskite vanadates, typically the family of $R\mathrm{VO_3}$ ($R$ being a rare earth or yttrium atom), possess highly tunable orbital and spin orderings with external stimuli \cite{phasediagram}. For example, in PrV$\mathrm{O_3}$ thin films, tunable magnetic properties have been achieved by epitaxial strain \cite{epitaxial1,epitaxial2}, or chemical strain through the control of oxygen vacancies \cite{chemical1,chemical2}. 
1 UC FeSe/SVO/STO, as a new realization of FeSe grown on the non-TiO$_x$ terminated substrate, inspires the future designing of the heterostructure in investigating strongly correlated physics, such as bandwidth control, relationship between superconductivity and magnetic orders, and electron-boson interactions.

\section{Method}

\noindent{\textbf{Sample fabrication.}} SVO films were epitaxially grown on the TiO$_x$ terminated SrTi$\mathrm{O_3}$ (001) substrates (Crystech GmbH) by using atomic layer-by-layer molecular beam epitaxy (MBE) method, at a substrate temperature of 750 $^{\circ}$C and an oxygen partial pressure of 1.9 $\times 10^{-7}$ mbar. The thickness of SVO was confirmed by the real-time intensity oscillation of reflective high-energy electron diffraction (RHEED) (Figure S1). FeSe films were grown on SVO subsequently at 490 $^{\circ}$C by co-evaporation method and annealed at 520 $^{\circ}$C for 3 h before ARPES measurement. ~\\
					
\noindent{\textbf{ARPES measurements.}} In-situ ARPES measurements were performed on FeSe/SVO/STO using a Scienta DA30 electron analyzer and a Fermi Instruments 21.2 eV helium discharge lamp, under a typical vacuum of 2 $\times 10^{-11}$ mbar. The total energy resolution is 6 meV and the angular resolution is 0.3 $^{\circ}$.~\\
					
\noindent{\textbf{HR-EELS measurements.}} For the in-situ HR-EELS measurements, the samples were annealed at $\sim$ 520 $^{\circ}$ in an ultra-high vacuum of 5 $\times$ 10$^{-9}$ mbar to get the clean surface. RHEED patterns were collected to verify the surface quality before HR-EELS measurements. The incident electron beam in HR-EELS measurements is produced by an electron gun of Model LK5000M (LK Technologies). The incident electron energy was 13.6 eV and the incident angle with respect to the surface normal was 45$^{\circ}$. Data were collected by an analyzer A-1 (MBS) under an ultra-high vacuum of 5 $\times$ 10$^{-11}$ mbar. The measurements were conducted at 6 K.~\\
					
\noindent{\textbf{STEM measurements.}} For STEM measurements, a 1.5 UC FeSe/SVO/STO sample was capped with Se and taken out of the vacuum chamber to protect the FeSe layers from being oxidized. Then the sample was prepared for STEM measurements by using Focused Ion Beam (FIB) milling. Cross-sectional lamella was thinned down to 100 nm at an accelerating voltage of 30 kV with a decreasing current from the maximum 2.5 nA, followed by fine polish at an accelerating voltage of 2 kV with a small current of 40 pA. The atomic structures of the FeSe/SVO/STO films were characterized using an ARM-200CF (JEOL, Tokyo, Japan) transmission electron microscope operated at 200 kV and equipped with double spherical aberration (Cs) correctors. HAADF images were acquired at acceptance angles of 90-370 mrad.~\\

\begin{acknowledgement}

This work is supported in part by the National Science Foundation of China under the grant Nos. 11922403, 12274085, 12074074, the National Key R\&D Program of the MOST of China (2023YFA1406300), the New Cornerstone Science Foundation, the Innovation Program for Quantum Science and Technology (Grant No. 2021ZD0302803), and Shanghai Municipal Science and Technology Major Project (Grant No.2019SHZDZX01). 

\end{acknowledgement}

\begin{suppinfo}

Details for sample preparation and experimental method, additional data from cross-sectional STEM.

\end{suppinfo}

\providecommand{\latin}[1]{#1}
\makeatletter
\providecommand{\doi}
{\begingroup\let\do\@makeother\dospecials
	\catcode`\{=1 \catcode`\}=2 \doi@aux}
\providecommand{\doi@aux}[1]{\endgroup\texttt{#1}}
\makeatother
\providecommand*\mcitethebibliography{\thebibliography}
\csname @ifundefined\endcsname{endmcitethebibliography}
{\let\endmcitethebibliography\endthebibliography}{}

\end{document}